\documentclass[prx,twocolumn,english,groupedaddress,floatfix,longbibliography]{revtex4-2}
\usepackage{amsmath,amssymb,amsfonts,graphicx,hyperref}

\hypersetup{
    colorlinks=true,
    linkcolor=blue,
    filecolor=magenta,      
    urlcolor=blue,
    citecolor=blue,
}

\begin{document}

\title{Transmon Phase Gates Controlled by Superconducting Soliton DAC}
\author{Derek Reitz}\author{Tony X. Zhou}\author{Aditya Sharma}\author{Ryan Bilotta}\author{John McFarland}\author{Aref Fouladi}\author{Jacob Glasby}\author{Aruna Ramanayaka}\author{Zachary Stegen}\author{Aaron Pesetski}\author{Mark Covington}\author{Gregory Boyd}\author{Jeremy Clark}
\affiliation{Northrop Grumman Corporation, Linthicum, Maryland 21090, USA}
\begin{abstract}
We introduce a superconducting digital-to-analog converter (DAC) that filters control noise, provides native multiplexing, performs quantum gates in nanoseconds, and can be controlled by CMOS. This is achieved by transducing a trapezoidal drive pulse into a superconducting soliton, which is then held in the DAC load loop, applying flux to a mutually-coupled superconducting qubit or gate coupler. The analog flux output by the DAC can be easily controlled by varying the soliton hold time, or with a DC-biased tunable DAC-qubit coupler, allowing the DAC to perform a fixed-time, high-fidelity gate that's robust to fabrication variance or flux offsets in the quantum circuit. Our initial demonstration shows that the DAC can successfully perform 5.6 ns S-gates on transmons. We measure the DAC-induced quantum state excitation probability per gate to be 0.05\%, and find that the DAC-induced relaxation rate from the qubit $|1\rangle$ state is below the intrinsic $T_1$ rate limit of the transmon. Quantum simulations show qualitative agreement with the measured data, and predict that the DAC excitation rate can be lowered 10 times further by overdamping the Josephson junction (JJ) in the DAC load loop. Interleaved Randomized Benchmarking (IRB) sequences on an observer qubit reveal that, when scaling to many qubits, the DAC's performance may be limited by a non-local, DAC-induced phase error of 1.6\% per gate, appearing in ancilla that are not directly coupled to any of the 30 DACs on the chip. We discuss strategies for future layouts of multi-DAC chips that focus on mitigating the source of these non-local, high-frequency electromagnetic interactions (EMI), and how to incorporate a DC-tunable coupler for phase correction.
\end{abstract}
\maketitle

\section{Introduction}
Traditionally, superconducting qubits have been directly controlled by room-temperature electronics that send finely-controlled waveforms down coaxial cables\cite{fowler2012surface}. More recently, single flux quantum (SFQ) pulses from triggered JJs have been used to perform quantum gates and anneal quantum circuits\cite{PhysRevApplied.2.014007, PhysRevApplied.11.014009, 9605311, 1211765, 9773260,  PRXQuantum.1.020314, geng, jiang, hu}. Cryo-CMOS\cite{9209175, 9895434} is also under development for quantum control applications. While quantum algorithms involve applying the same gate to a patch of physical qubits\cite{fowler2012surface}, each physical quantum gate will require a slightly different control waveform due to fabrication variance and trapped flux offsets. Scaling up room-temperature or SFQ control thus encounters a wiring bottleneck\cite{markov2014limits, bernhardt2025quantum}, from the need to individually tune qubit control with a fixed number of wires. Requiring an individual room-temperature microwave line for each qubit is not only problematic for scaling in the long term, but may result in significant heating of the qubit environment for near- or intermediate-term applications. On the other hand, a cryo-CMOS control chip that's bump-bonded to the quantum chip could feedback information to modify its control signals with more limited room-temperature wiring\cite{gray2009analysis, xue2021cmos, PhysRevApplied.13.054072}. Here, the soliton DAC also plays a critical role in noise buffering, since the cryo-CMOS chip will likely operate near 10K due to self-heating\cite{hart2021characterization}, which would decohere a directly-coupled superconducting qubit. \par
The soliton DAC can be driven by a cryo-CMOS source, and thus provides a solution to both the wiring bottleneck and control noise filtration issues. We taped out a chip with 30 soliton DACs, and demonstrated that a single noisy control line can be used to drive many DACs on a single chip without degradation in the passive $T_1$ and $T_2^*$ times of the qubits they control. This approach can be scaled to large numbers of individually-tunable quantum gates by including a DC-biased tunable coupler that controls the coupling between the DAC and the qubit. This effectively trades a microwave line for a DC line. This DC bias can then be multiplexed with a superconducting flux-memory device\cite{dayton2018experimental}, which has been demonstrated to be a robust, scalable technology\cite{6802426}. Our novel control architecture proposal involves combining flux memory with the soliton DAC, forming the basic ingredients necessary for enabling scalable, noise-free, analog control of superconducting circuits in a way that's compatible with generic, digital input. Here, the superconducting DAC serves as an active buffer between a classical control system (whether its cryo-CMOS, low-pass filtered room-temperature CMOS, or another source) and bump-bonded superconducting quantum circuits\cite{PRXQuantum.4.030310}. Both the soliton DACs and flux memory can be co-located in a cryogenic environment with the quantum circuit. The soliton DAC has wide applicability, since there are many flux-controllable qubits, such as transmons\cite{PhysRevA.76.042319}, flux qubits\cite{orlando1999superconducting}, and fluxonium\cite{manucharyan2009fluxonium}, and many flux-controllable gate couplers\cite{PhysRevApplied.19.044031, ding2025pulse, glaser2024sensitivity, 10.1063/5.0304764, negirneac2021high}.\par

In this paper, we demonstrate the design, test, and simulation of a proof-of-principle CMOS-controlled superconducting soliton DAC where a current bias launches a soliton to perform an S-gate. The S-gate involves applying a frequency shift $\delta f_{01}$ between the qubit $|0\rangle$ and  $|1\rangle$ states until the relative phase accumulated reaches $\pi/2$. The S-gate can be performed by first DC-biasing the transmon, and then applying an external time-dependent flux to the transmon loop. Our DAC generates this packet by injecting a soliton when the input current exceeds a critical threshold ($\mathrm{I_{\mathrm{crit}}}$), holding for 5.6 ns, and then removing the bias by injecting an anti-soliton after the input bias is turned off. As the input bias amplitude increases, the DAC fires at a progressively earlier point in time, generating a hallmark interference pattern that's seen in both simulation and test.

\section{Background}
\begin{figure}[htbp]
\centerline{\includegraphics[width=0.5\textwidth]{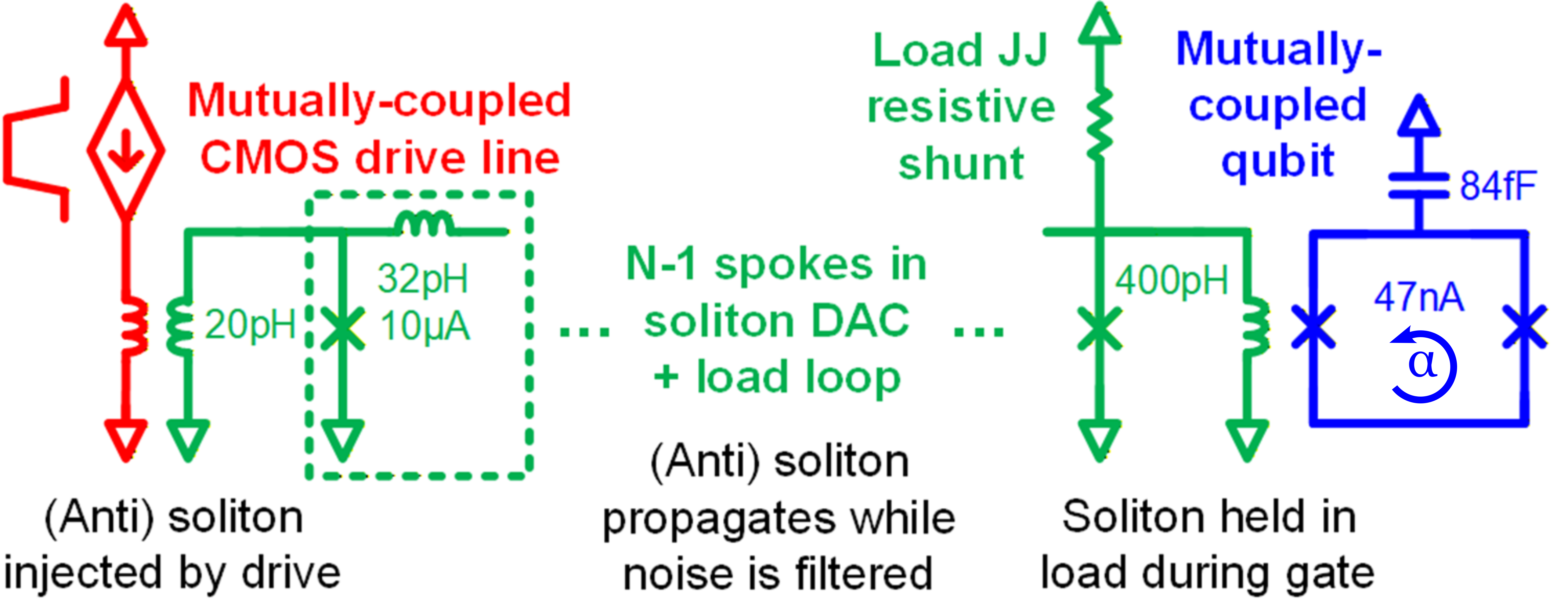}}
\caption{Schematic showing the soliton DAC with mutually-coupled drive line and mutually-coupled transmon target loop. Each spoke, containing a grounded JJ and spine inductor, forms the tile of a repeating lattice in the middle of the DAC. Resistors can be added to shunt each of the JJs, but the speed of the flux waveform that's applied to the target is largely controlled by the resistive shunting on the final JJ, shown above.}
\label{fig1_sim}
\end{figure}

The soliton DAC is an inductively-coupled chain of grounded JJs, terminating in a load loop where flux can be thrown to a target loop. We focus on the design shown in Fig.~\ref{fig1_sim}, with a damping resistor on the final JJ to smooth the square-wave output waveform. The soliton that propagates the signal from the input drive down the Josephson Transmission Line (JTL) is well-described in Mclaughlin and Scott's paper\cite{mclaughlin1978perturbation}. The discretized JTL is the Sine-Gordon lattice, where flux in a superconducting loop and Kirchhoff's Current Law require that,
\begin{equation}
\frac{\Phi_0}{2\pi}\left(\varphi_{n}-\varphi_{n+1}\right) = L_sI_n
\end{equation}
and
\begin{equation}
I_{n-1}-I_n = \left(C\partial_t^2 + G \partial_t \right) \varphi_n + I_c \sin\varphi_n,
\end{equation}
respectively, for current $I_n$ through the nth inductor with spine inductance $L_s$, superconducting phase $\varphi_{n}$ across the nth JJ with critical current $I_c$, JJ capacitance $C$ and JJ shunt resistance $R = 1/G$ (including the subgap resistance). In the continuum limit, combining these two equations yields the Sine-Gordon equation,
\begin{equation}
J/J_c = \left( \partial_{\tau}^2  - \partial_X^2 + \alpha \partial_{\tau} \right)\phi + \sin\phi,
\label{sge}
\end{equation}
where $J$ is the applied current density, $J_c$ is the critical current density, $\alpha = \sqrt{\Phi_0 / 2 \pi J_c R^2}$ is the damping constant, and we have switched to natural units $\tau = t f_0$ for $f_0 = \sqrt{2 \pi I_c / \Phi_0 C}$, and $X = x / \lambda_0$ for $\lambda_0 = \sqrt{\Phi_0 / 2 \pi J_c L_s}$. Compared to a long JJ, here, the capacitance, inductance, critical current, and resistance are all tunable. The undamped, undriven limit of Eq.~\ref{sge} admits soliton solutions of the form:
\begin{equation}
\phi(X, \tau)=4 \tan^{-1} \exp\left(\pm\frac{X-v\tau}{\sqrt{1-(v/c)^2}}\right),
\end{equation}
for soliton velocity $v$ and plasma velocity $c = \lambda_0 f_0$. The kink-antikink solutions describe a rising or falling edge of the soliton in the JTL. Here, the + corresponds to a soliton and - to an anti-soliton; or, equivalently, a soliton moving to the right or to the left.\par

At the time, the JTL was recognized for its central role in the ``transmission, storage, and processing of information'' in superconducting electronics\cite{mclaughlin1978perturbation}. The JTL may now be useful for superconducting quantum computing since it can translate a digital input into a fast flux bias to quantum circuits and filters the input signal noise. This filtration effect stems from the fact that noise amplitudes below the JJ critical current simply pass through to ground, with the JTL's noise suppression increasing with the number of spokes, by current division. At the same time, the Sine-Gordon soliton propagates down the JTL without degradation by triggering each JJ it passes over, resulting in an output flux with an enhanced signal-to-noise ratio.\par

The essential requirements for a functional DAC are: repeatable injection and removal of solitons, which is satisfied when 1) the spoke $\beta \equiv L_s / L_{JJ}$, for Josephson inductance $L_{JJ}=\Phi_0/2 \pi I_c$, is in the range $0.3 < \beta < 0.9$, 2) stable retention of solitons in the load loop, which is satisfied in WRS for load $\beta_{load} \equiv L_{load} / L_{JJ} > 1.2$, and 3) there are at least as many spokes as the length of the soliton. These three requirements are verified by WRSpice (WRS) simulation\cite{wrspicemanual}. As pointed out in Ref.~\cite{mclaughlin1978perturbation}, fabrication variation of the critical currents translates into a random potential landscape, which acts as a force in the solitonic equation of motion. We reproduced this effect in WRS Monte Carlo simulations, and found that, with $I_c$ disorder, a drive current that only exceeds the soliton injection threshold by 1\% can result in trapping of the soliton in the middle of the JTL. This results in zero output at the load, since the soliton remains trapped until it's annihilated by the anti-soliton that's injected when the drive is turned off. However, we also found that the pinning potentials generated by $I_c$ disorder can be overcome with a modest 10\% increase in the drive amplitude above the soliton injection threshold. Furthermore, a small negative DC bias, with amplitude equal to 10\% of the injection threshold, avoids anti-soliton trapping. When a small negative DC bias was used in combination with the drive pulse, our DACs did not show trapping, latching, or other hysteretic complications during test. The soliton DACs that we fabricated had $\mathrm{I_{\mathrm{crit}}} = $ 340$\mu$A, with a 6 pH input mutual.

\section{Design and passive filtration}
\begin{figure}[htbp]
\centerline{\includegraphics[width=0.5\textwidth]{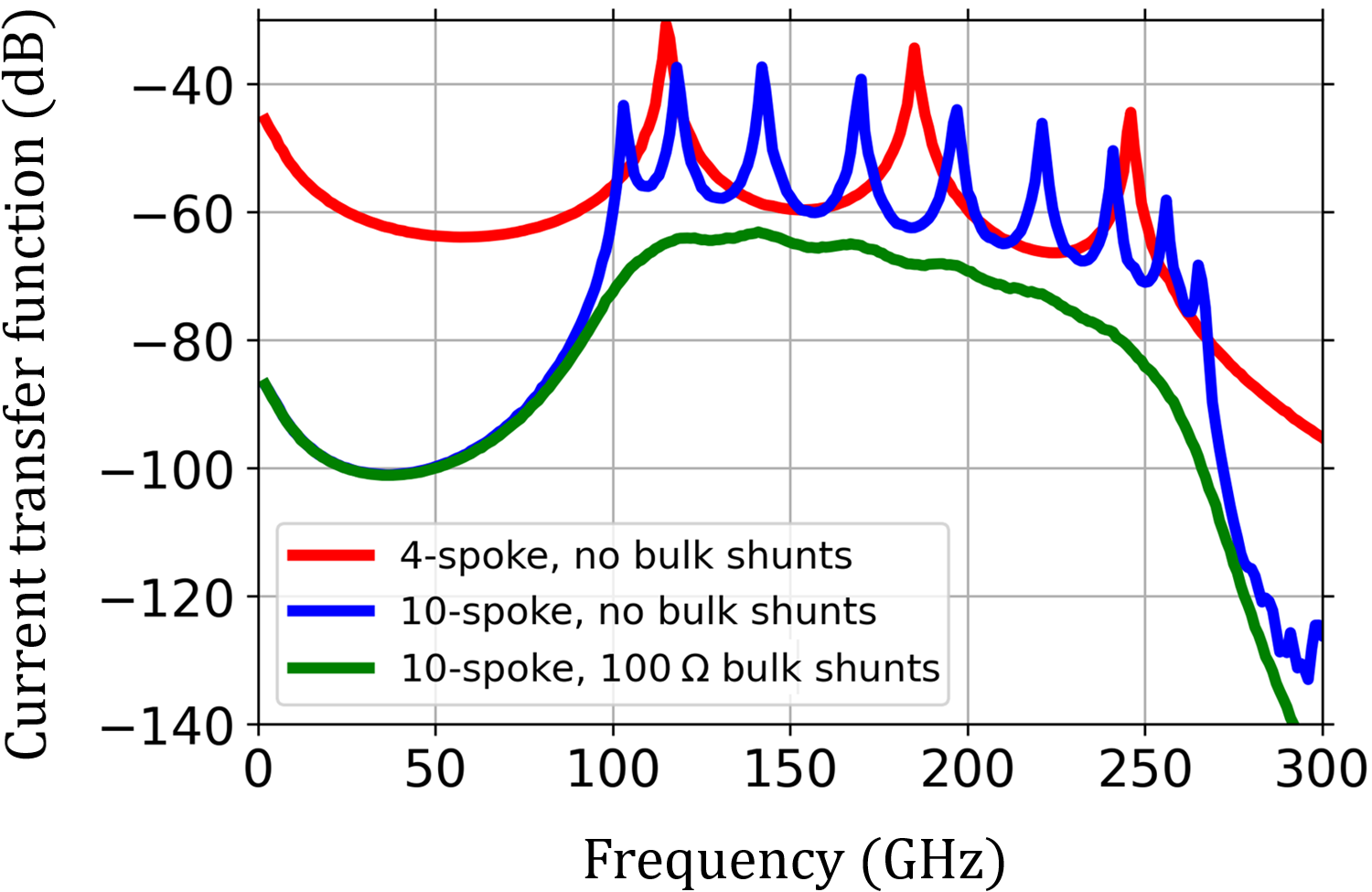}}
\caption{WRSpice-simulated square of the current transfer function, relating the current in the DAC's final load to first input inductor, which is independent of the mutual coupling strengths. The noise filtering capability of the DAC is demonstrated by the relatively smaller output noise amplitude, in the qubit frequency range, with ten JJs (green, blue) compared to four JJs (red). The JJ plasma frequency, $f_J = 267$GHz, manifests as high-frequency peaks associated with plasma modes that can be excited within the DAC. The resonances can be damped out by including 100 $\Omega$ shunts on every JJ in the middle of the DAC (green).}
\label{fig2_sim}
\end{figure}

The DAC we designed and tested is shown in Fig.~\ref{fig1_sim} with 10 DAC spokes, and the size of the soliton in WRS simulation was approximately 1.5 JJs. The continuum model becomes exact in the limit where $\beta\ll1$. Our 10-spoke soliton DAC has $\beta$=0.60, and we found that underlying physics seen in WRSpice simulation was essentially described by a continuum model for the JTL, Eq.~\ref{sge}. The number of spokes was chosen based on a tradeoff between the physical size of each DAC on the chip and its noise filtration capability, which is shown for several designs in Fig.~\ref{fig2_sim}. The 10-spoke, 100 $\Omega$ shunted design shows -60 dB or better attenuation across 300 GHz of bandwidth. Transmon $f_{01}$ frequencies are typically 5-12 GHz, where the attenuation is better than -80 dB. Therefore, the soliton DAC provides native multiplexing, with a below threshold pulse attenuation factor of order $10^3$. If a greater level of reduction is needed for applications, additional multiplexers can be attached to the input-side of the soliton DAC to act on top of the native low-amplitude pulse rejection capability. For a N-spoke JTL, there are N-1 resonance peaks in the noise transfer function associated with plasma modes, which can be damped out with 50-100 $\Omega$ shunts on each JJ in the middle of the DAC. WRS simulation of the DAC confirms that, with a 10-spoke, $\beta=0.5$ JTL, $R > 10$ $\Omega$ bulk shunts introduce a negligible, picosecond delay between the input and output waveforms. 
The experimental apparatus design that we used is shown in Fig.~6 of Ref.~\cite{PRXQuantum.1.020314}. The reference microwave gates are performed by a separate microwave drive. The Z rotations are achieved by applying perturbative, variable-height trapezoidal waveforms of approximately 5ns duration to the transmon $\alpha$ loop. The microwave Z-rotation pulse duration was approximately equal to the DAC pulse. XY-gates are performed on the transmons via a radio frequency tone that's applied to a capacitively-coupled drive line. The tone is carried through a signal line designed in the cryostat, and then onto the chip package. The red drive line in Fig.~\ref{fig1_sim} is shared across all DACs on the chip. In addition, the chip contained a set of ancilla qubits that are not connected to the DAC drive line. 

\begin{figure}[htbp]
\centerline{\includegraphics[width=0.5\textwidth]{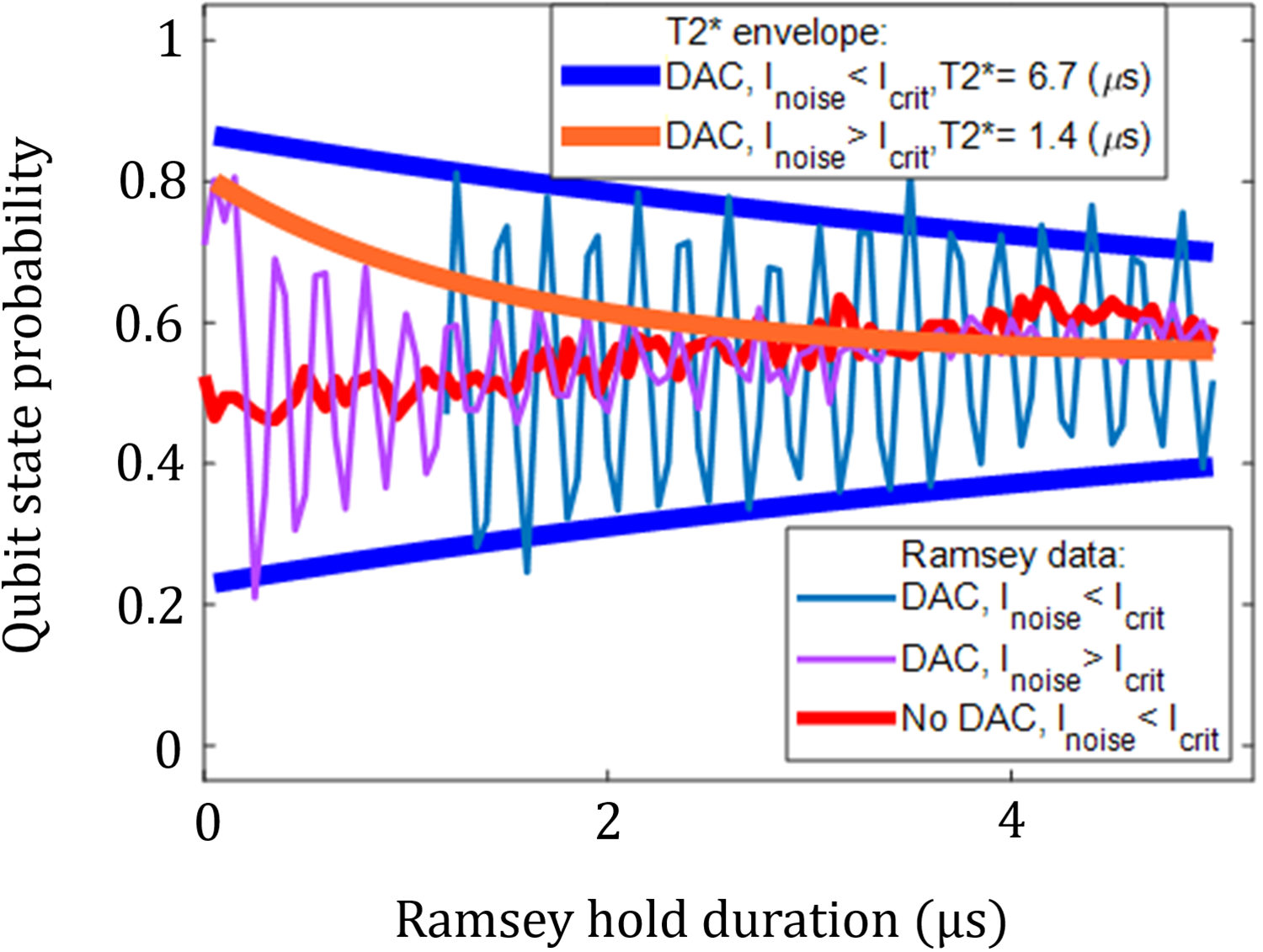}}
\caption{Soliton DAC qubit protection demonstration via Ramsey measurements of a transmon at its upper sweet spot. Control noise, with amplitude $\mathrm{I_{\mathrm{noise}}}$, is injected into the transmon’s $\alpha$ loop. When 1V peak-to-peak noise is injected through conventional refrigerator attenuators and a 780 MHz cutoff-frequency, passive, low-pass filter into an ancilla qubit (red trace)–that’s not protected by a DAC–the Ramsey fringes are not detectable, resulting in complete qubit degradation. In contrast, the soliton DAC protects the transmon from 1 V noise to retain its noise-free $T_2^* \approx 6.7$ $\mu$s coherence (teal trace and blue curve envelope fit). When the noise amplitude exceeds the soliton injection threshold with 4 V noise, we observe slightly attenuated Ramsey oscillations corresponding to $T_2^* \approx 1.4$ $\mu$s (purple trace and orange curve envelope fit).}
\label{fig1_exp}
\end{figure}

The theoretical, perturbative-noise filtration discussed above was validated in test by measuring the effect of control noise on $T_2^*$ of the transmons. In Fig.~\ref{fig1_exp}, we compare the results for a Ramsey experiment with control noise either applied directly to the transmon's $\alpha$ loop (see Fig.~\ref{fig1_exp}), or applied through the soliton DAC. We observed no measurable degradation of  $T_2^*$ until the control noise amplitude, $\mathrm{I_{\mathrm{noise}}}$, exceeded the soliton injection threshold, $\mathrm{I_{\mathrm{crit}}}$. Since $\mathrm{I_{\mathrm{crit}}}$ is the pulse amplitude needed to perform a quantum gate, any noisy drive with a reasonable signal-to-noise ratio that's greater than $\sim10$ would be capable of driving the DAC without degrading transmon $T_2^*$. For a given $\mathrm{I_{\mathrm{noise}}}$, the flux noise power applied to the Transmon through the direct path is 0.14 times the noise applied through the DAC path, due to an additional mutual coupling. That being said, since we see complete $T_2^*$ degradation (when the qubit is not protected by the DAC) for even small $\mathrm{I_{\mathrm{noise}}} \ll \mathrm{I_{\mathrm{crit}}}$, we conclude that the DAC is successfully filtering control noise below the threshold that's required for quantum computation.

For the qubit, we used a standard transmon with target critical current of $I_{c, Tmon} = 47$ nA, shunt capacitance $C_{Tmon}= 83$ fF, and DAC-qubit mutual coupling of $M = 4.6$ pH, depicted in Fig.~\ref{fig1_sim}. The transmons’ anharmonicity was in the typical range to render the Transmon insensitive to charge fluctuations. To generate the control noise, we used the Function Generator Agilent 33220A and its DC noise function to output 9MHz bandwidth random voltage fluctuations with an adjustable peak to peak voltage. The transmons were readout with flux qubits, using the method described in Ref.~\cite{abraham2026fast}.

\section{Gate performance}
Two superconducting chips were fabricated and tested at Northrop Grumman's Advanced Technology Laboratory. The control chip contained 30 superconducting soliton DACs, with the load of each DAC mutually coupled to a corresponding transmon on the superconducting qubit chip. Tune-up of the transmon phase gate involved measuring the transmon spectrum, preparing and reading out the transmon qubit states, and measuring the change in qubit phase vs. the amplitude and hold time of the trapezoidal waveform that drives the DAC. The performance of the DAC was characterized by DC-biasing the transmon to the point where $d f_{01}/d \Phi$ = 4GHz / $\Phi_0$ and transmon $f_{01} = 6.8$ GHz, and then driving a single soliton into the DAC load loop with a gate time $t_{\mathrm{gate}} = 5.6$ns. The DAC load loop is mutually coupled to the transmon $\alpha$ loop, so that the total flux $\delta \Phi$ thrown by the DAC results in a $90^\circ$ phase: $\theta = \pi/2 \approx 2\pi t_{\mathrm{gate}} \; \delta \Phi d f_{01}/d \Phi$. Two 10-spoke designs were tested, one with a single 10 $\Omega$ damping resistor at the end of the DAC, and one with three 10 $\Omega$ resistors on the final three JJs of the DAC. The three-shunt design had slightly higher gate fidelity, so we focus on that design below.\par

\begin{figure}[htbp]
\centerline{\includegraphics[width=0.5\textwidth]{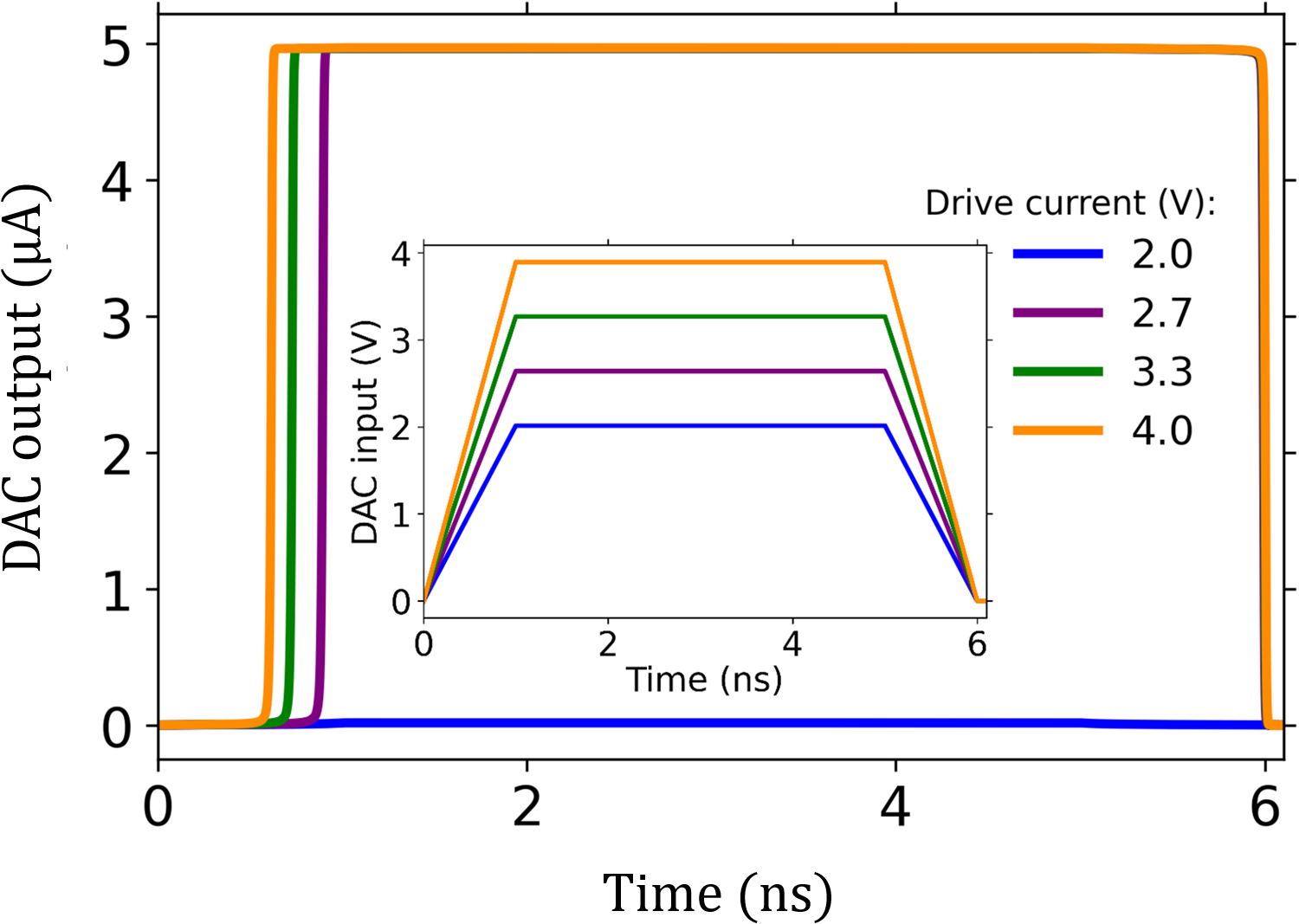}}
\caption{Four idealized, trapezoidal waveforms with varied amplitudes drive the soliton DAC in a WRSpice simulation. The lowest amplitude drive (blue) does not exceed the soliton threshold, and results in a simulated output that's $10^3$ times smaller than when injection occurs. As the drive amplitude is raised above the threshold, but remains below twice the injection threshold (where a second soliton is injected) the output amplitude remains constant. The flux thrown by the DAC, however, decreases slowly as the drive amplitude increases above threshold. This is because the trapezoidal drive passes the injection threshold earlier, triggering the rising edge of the output earlier, while the falling edge remains fixed at the point where the output waveform reaches zero.}
\label{fig3_sim}
\end{figure}

\begin{figure}[htbp]
\centerline{\includegraphics[width=0.5\textwidth]{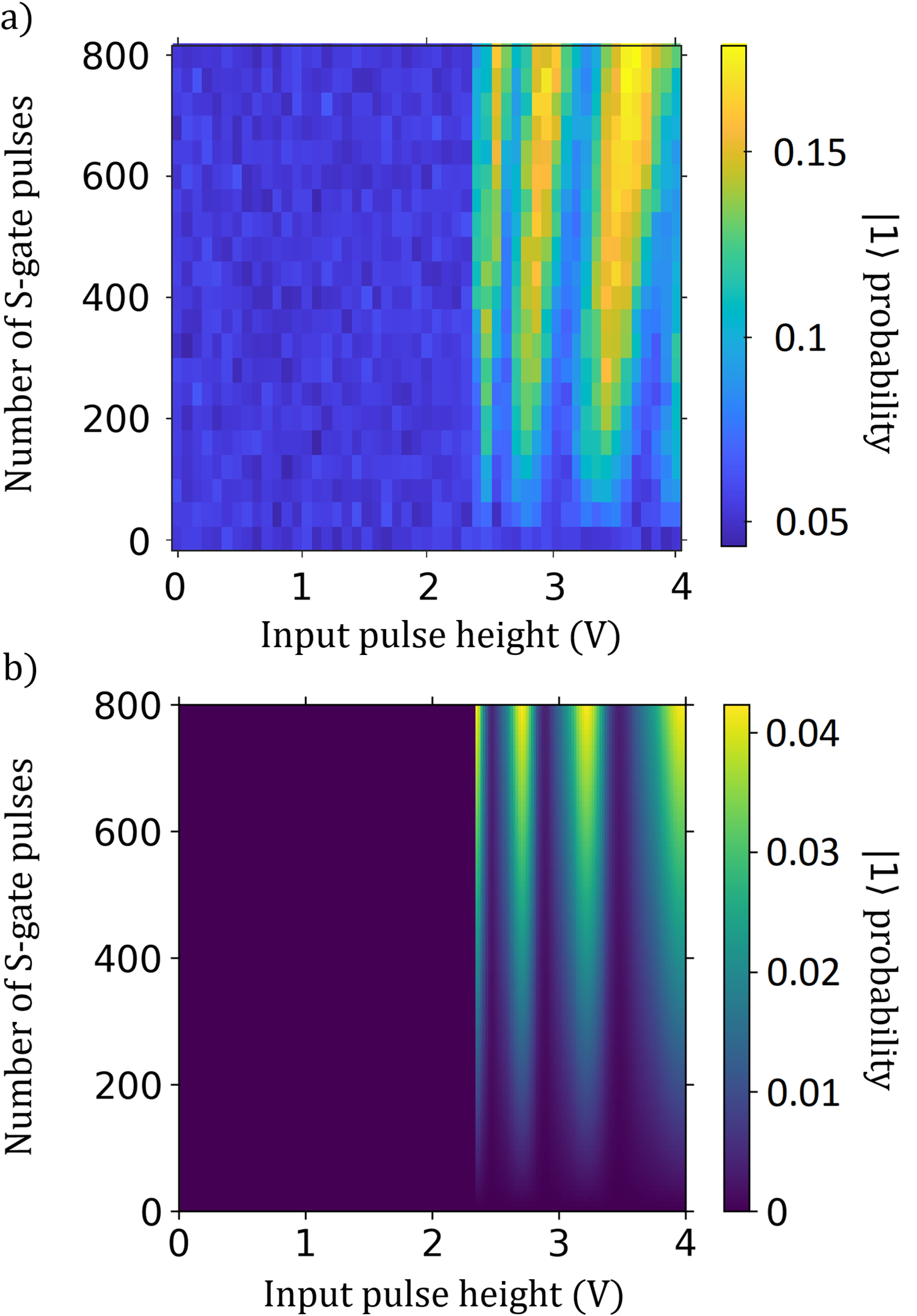}}
\caption{The transmon is prepared in the $|0\rangle$ state and excited by the rising and falling edges of the DAC waveform, with the final $|1\rangle$ probability plotted for test data, a), and simulation, b). The soliton DAC waveforms that perform transmon S-gates were generated by WRS and fed into a Circuitizer simulation of the quantum circuit. The excitation pattern is a result of a coherent oscillation between the initial and excited states, where the oscillation duration changes with the bias amplitude, as shown in Fig.~\ref{fig3_sim}. Compared to the experiment, the simulated excitation probability is lower, in part due to the fact that the physical layout contained 30 soliton DACs which are simultaneously triggered, driving non-local errors from EMI.}
\label{fig4_sim}
\end{figure}

In order to investigate the qubit excitations generated by the DAC, we measured the $|0\rangle \rightarrow |1\rangle$ transition probability vs. the number, $N$, of DAC S-gate pulses and the amplitude of the input waveform driving the DAC. We observe an oscillatory excitation pattern vs. the DAC drive amplitude that is consistent with simulation. The DAC rejected the input trapezoidal pulse when its amplitude was below the soliton injection threshold, resulting in zero measurable DAC-induced relaxation after 800 pulses, relative to the prepared state. This confirms the capability of the DAC to multiplex the input, for example, in a crossbar arrangement\cite{6802426}. As the amplitude increases above the threshold, the slope of the trapezoidal drive increases, so we inject the soliton earlier, as shown in Fig.~\ref{fig3_sim}. Soliton injection is associated with a rapid change in flux in the transmon, generating qubit excitations. In the experiment, the delay time between successive S-gates was kept constant. Thus, the physics of the excitation oscillation pattern shown in Fig.~\ref{fig4_sim} is the same as an interferometer--the peaks and dips come from constructive and destructive interference between the state that was prepared before the DAC was pulsed, and the state that we leaked into after the DAC was pulsed. For total times elapsed during pulsing the DAC that are comparable or longer to the measured transmon $T_1=8 \pm 0.3$ $\mu$s, the measured $|1\rangle$ probability is affected by decoherence; however, for short elapsed times compared to $T_1$, the $|1\rangle$ probability increases linearly in $N$, yielding an average excitation probability per gate of 0.05\% over the first 100 gates at the peak (maximal destructive interference). We quote the maximal excitation probability per gate, since the timing of the read operation relative to the phase gate may vary in applications. Note that the data in Fig.~\ref{fig4_sim} demonstrates that the injected soliton is removed each time the DAC input drive is turned off. This is because qubit excitations only occur when the flux in the transmon is changing over time. In turn, this only occurs when a soliton is injected and removed from the DAC load loop, and the measured $|1\rangle$ probability increases with the number of DAC pulses.

\begin{figure}[htbp]
\centerline{\includegraphics[width=0.5\textwidth]{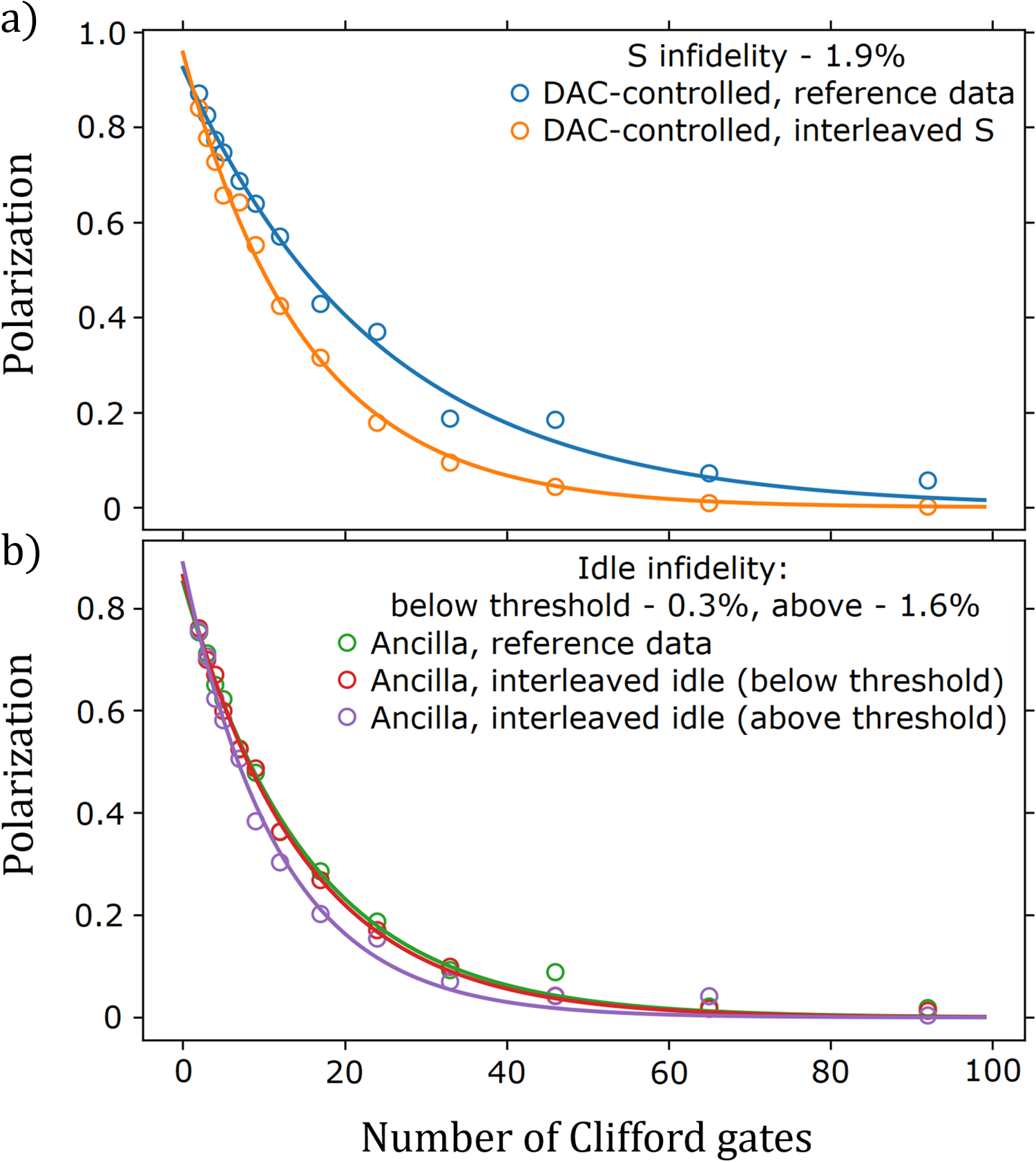}}
\caption{Measured Character IRB data for a) soliton DAC-controlled S-gates and b) idle gates while the soliton DAC input control line is pulsed. When the DAC input drive amplitudes are below the soliton injection threshold, the idle gates are indistinguishable from the reference microwave gates. On the other hand, when the drive amplitude exceeds the soliton injection threshold, an ancilla idle error is observed that's within our measurement precision of the DAC S-gate error. The solid lines are fits.}
\label{fig2_exp}
\end{figure}

The IRB data for a 5.6 ns soliton DAC S-gate is shown in Fig.~\ref{fig2_exp}a\cite{magesan, crb}. The gate error can be attributed to four sources: unwanted qubit transitions due to GHz-frequency content in the rising and falling edges of the DAC waveforms and the DAC admittance, high-frequency phase error due to jitter in the relative timing of the rising and falling edges, non-Markovian phase errors, and constant ``tune-up'' imprecision in the overall phase thrown by the DAC. Importantly, most of the error associated with pulsing the DACs was observed to come from triggering DACs that are not directly connected to the measured qubit. We quantified this non-local error by performing IRB on ancilla qubits with no direct connection to any of the DACs, shown in Fig.~\ref{fig2_exp}b. In the ancilla, when the DAC input pulse amplitude was below their trigger threshold, we measured an idle IRB error (over the same S-gate time) of $0.25\pm0.49\%$, compared to $1.6\pm0.60\%$ when it was above the threshold. Using a measurement sequence that's similar to the one that generated the data shown in Fig.~\ref{fig4_sim}, except that the qubit was prepared in the $|1\rangle$ state instead, we measured the relaxation rate versus the DAC input pulse amplitude and observed no dependence. Therefore, the qubit relaxation rate, $\gamma_{\downarrow}$, was determined purely by the intrinsic decoherence of the transmons. The DAC-induced excitation rate, $\gamma_{\uparrow}$, in Fig.~\ref{fig4_sim}a, was 36 times smaller than the DAC gate errors, so this was also a minor contribution. Our AC control lines were filtered with 780 MHz low-pass passive filters. Similar to the IRB fidelity estimates, IRB uncertainties were estimated via a binomially shot-weighted linear least-squares (chi-squared). State preparation and measure error, which can bias this estimate, was small enough to be negligible.

Finally, we performed a Ramsey experiment on a DAC-connected qubit to characterize the phase errors. Here, the qubit was prepared on the equator of the Bloch sphere and the DAC performed a variable number of T-gates. Compared to the DAC S-gate IRB infidelity of $1.9\pm 0.4\%$, we measured a DAC T-gate infidelity of $1.8\pm 0.7\%$, showing that the gate errors were independent of the DAC hold time. Rather, the data indicates that they are generated from the rising and falling edges of the DAC. The fidelity of the T-gate was estimated from analysis of the Ramsey and DAC-induced excitation/relaxation experiments, and it agrees with both the fidelity of the DAC S and ancilla idle with DAC drive amplitude above threshold, to within error bars. This indicates that the dominant source of the phase noise is independent of the hold time and originates from activation of the DAC. Simulations indicate that the phase noise may originate in the underdamped JJs in the middle of the DACs, which must be coupling in to the transmons directly, based on our ancilla data. Since all 30 DACs on the chip have the same JJ critical currents and share the same drive line, they are all simultaneously triggered when the drive line is pulsed. Therefore, we hypothesize that this error may come from non-local excitations driven by JJs in the middle of each DAC firing simultaneously. Based on these results, we suggest future chips contain a smaller subset of DACs, that are controlled by their own drive line, in order to assess the qubit errors vs. the number of DACs that fire simultaneously. Furthermore, we propose adding resistive shunting to the JJs in the middle of the DACs to potentially mitigate this effect. \par
Frequency content above 20GHz does not directly affect the qubits in a quantum simulation, but it can have indirect effects. Specifically, quasiparticles in the transmon JJs can be generated by the high-frequency EMI, which can cause heating of the qubits. During the quasiparticle relaxation process, lower, GHz-frequency photons can be emitted which directly drive qubit transitions\cite{PhysRevB.96.104510}. High-frequency radiation by the DAC is mitigated, in a WRSpice simulation, by adding bulk resistive shunting, as shown in Fig.~\ref{fig2_sim}. 

Our setup was more fundamentally limited by the 200ps hold-time resolution of the Tektronix 5208 Arbitrary Waveform Generator (AWG) that drove the DAC, which would result in a lower bound for the S-gate phase error of approximately 0.1\%. Once the long-range electromagnetic interactions are mitigated, in order to reach the maximal simulated gate fidelity, one would either need to improve the hold-time precision of the driving waveform by 10 times to approximately 20 ps for a 5 ns S-gate, or attach a tunable coupler between the DAC and the qubit. The DC bias to the tunable coupler could then be multiplexed, as discussed in Ref.~\cite{6802426}, to form a scalable superconducting control architecture. The device we designed had a simulated DAC slew rate of 10 ps and simulated DAC-induced excitation probability per gate of $5 \times 10^{-5}$, shown in Fig.~\ref{fig4_sim}b. While the simulation only contained a single DAC controlling the transmon, the chip contained 30 DACs which were triggered simultaneously by their shared drive line. As discussed above, we observed significant non-local errors due to the other DACs on the chip. Modeling these effects would require accounting for how EMI in the 1-300GHz range propagates across the chip and affect the qubits, which was beyond the scope of our initial demonstration. \par

\begin{figure}[htbp]
\centerline{\includegraphics[width=0.5\textwidth]{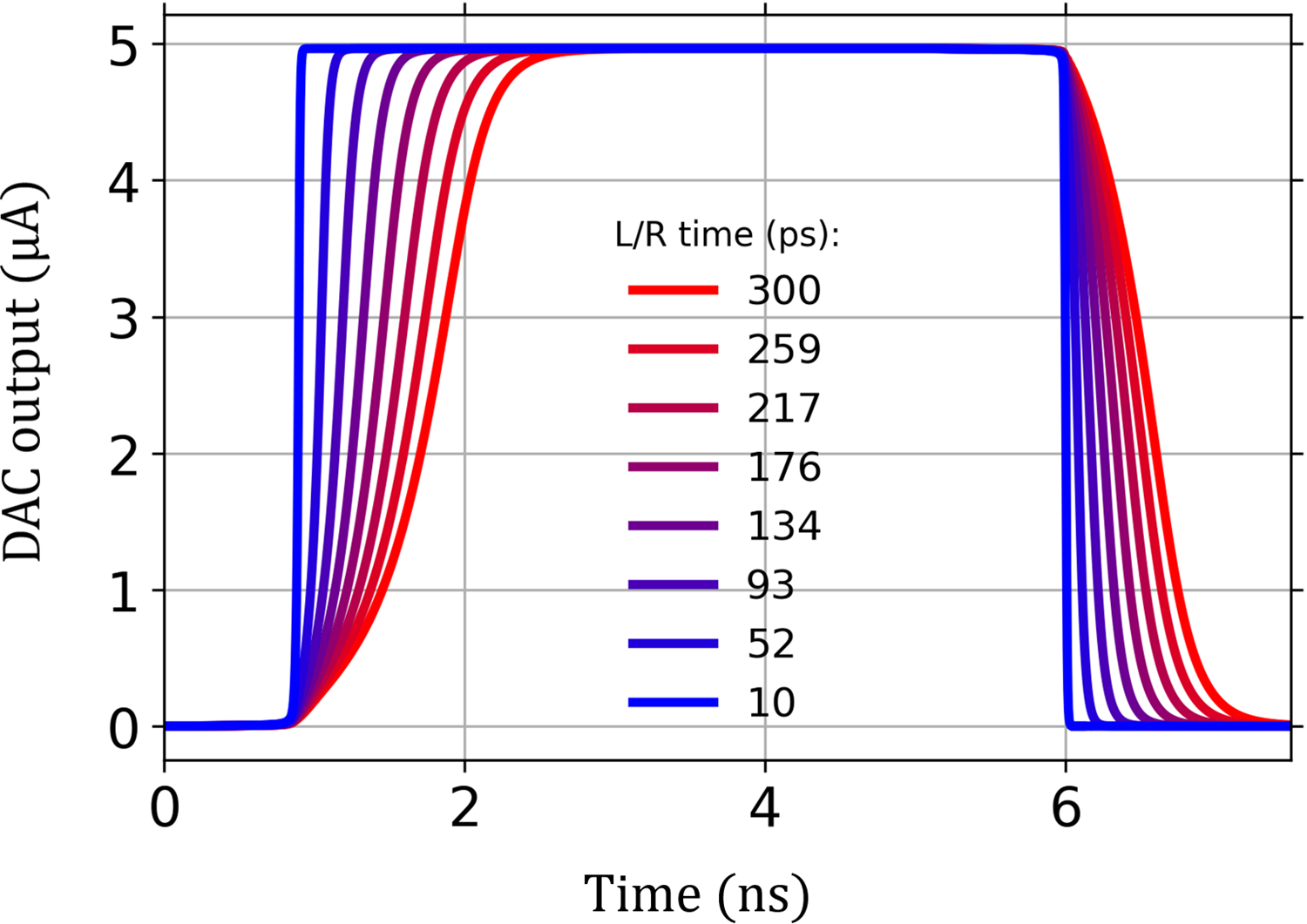}}
\caption{WRSpice simulation of the DAC waveforms for varied shunt resistance of the final JJ, given a fixed input waveform. The DAC variants are labeled by their equally-spaced $L/R$ times, where $L$ is in the inductance of the DAC load loop. The output waveform slew time is correspondingly equally-spaced, showing that it is determined by $L/R$.}
\label{fig5_sim}
\end{figure}

While the solitons propagate ballistically in the JTL at the plasma velocity, the DAC output waveform slew rate (i.e., trapezoidal rise and fall time) is instead determined by the $L/R$ time of the load loop, where $L=(L_{load}^{-1}+L_{JJ}^{-1} )^{-1}$. The simulated waveforms for our 10-spoke DAC with varied final shunt resistance and a fixed driving waveform are shown in Fig.~\ref{fig5_sim}. Waveforms with faster slew rates have relatively higher PSD content at the qubit transition frequencies. The qubit excitations generated by these waveforms are discussed below. 

\begin{figure}[htbp]
\centerline{\includegraphics[width=0.5\textwidth]{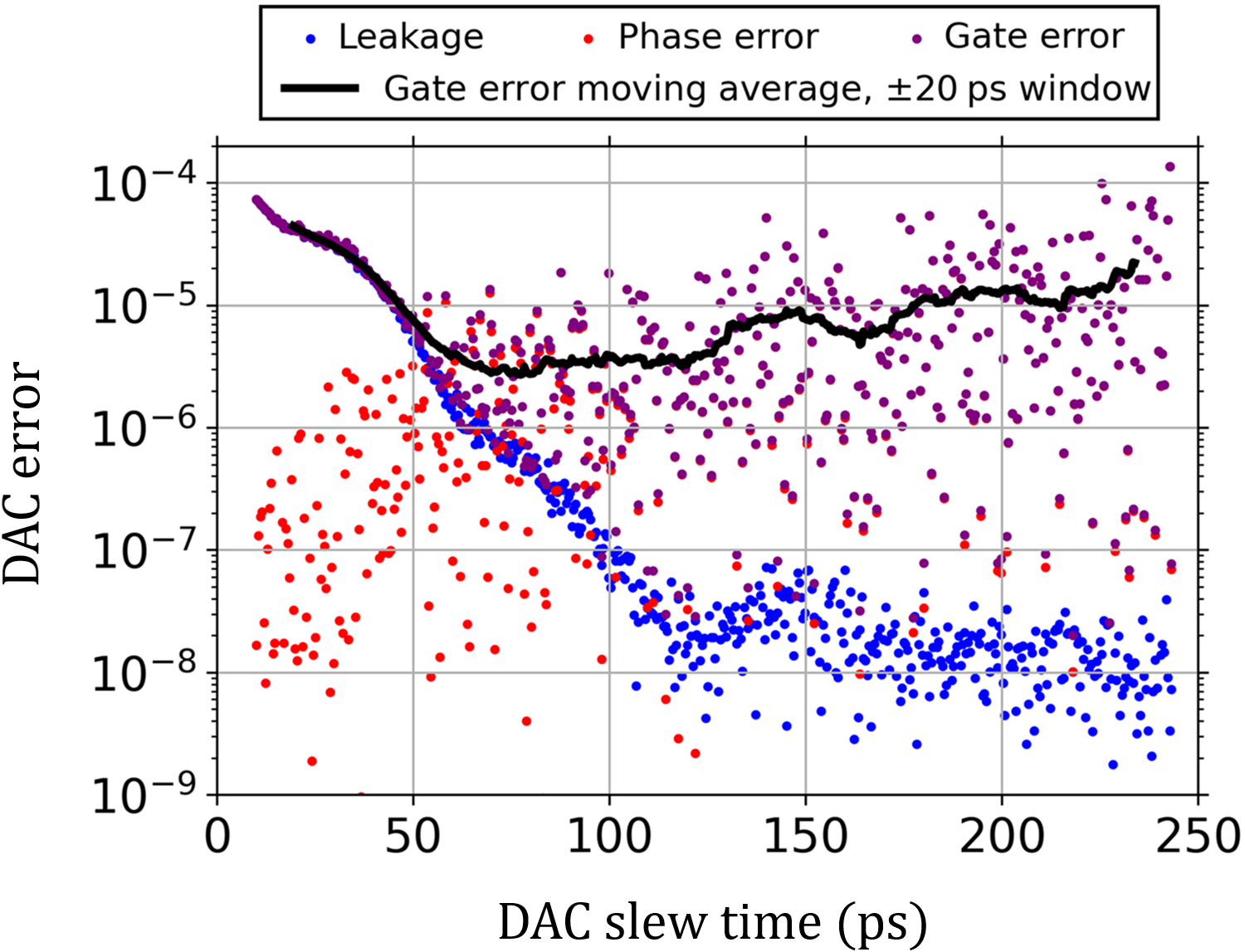}}
\caption{500 Circuitizer simulations for varied DAC load loop $L/R$, or waveform slew, times. Each simulation contained a randomly-sampled time trace of the Johnson-Nyquist noise that's emitted by the resistor shunting the final DAC JJ. Since the waveforms vary smoothly with $L/R$, we average over a sliding window for estimating the average gate error probability, shown in black. As the DAC waveform slew rate is progressively slowed, the resistor noise-induced phase errors (red) eventually overcome the savings in DAC-induced leakage errors (blue), resulting in a minimum in the gate error (purple).}
\label{fig6_sim}
\end{figure}

The DAC-induced error was simulated using WRS-generated waveforms as input to Circuitizer, Northrop Grumman's superconducting circuit quantum simulator, including the Johnson-Nyquist noise from the resistive damping of the final DAC JJ. Each data point in Fig.~\ref{fig6_sim} represents the error probability from one “shot” of noise that’s added into the DAC simulation with a time-dependent current source, implementing the Norton-equivalent resistor circuit. The current source is a sample of the noise PSD over the gate time, with the standard method described in Ref.~\cite{timmer1995generating}. Here, we assumed a nominal DAC temperature of $T$=0.1K, and used the frequency-symmetrized Johnson-Nyquist PSD $P(f) = 2k_B T n(f) / R$, for Bose-Einstein distribution function $n(f)$. The PSD is symmetrized in the frequency domain in order to generate a real-valued sample time trace of the PSD. This trace can then be applied as the Norton-equivalent circuit noise source in WRSpice for the DAC’s resistor.

When considering a DAC design, one free design parameter is the mutual coupling to the quantum circuit. In each simulation, we selected the mutual based on the noise-free flux waveform output by the DAC. As we vary the DAC's damping resistance, and therefore the DAC waveform slew rate, there is a small change in the net flux applied to the transmon. We set the mutual such that the noise-free phase applied by the DAC is $\pi/2$. In design, a DC-controlled tunable coupler would replace the fixed mutual that was used in our demonstration. Finally, the DAC S-gate infidelity was calculated by leakage + $\theta^2/4$, where leakage is defined as the net probability to leave the ideally-prepared $|1\rangle$ state, and $\theta$ is the angle difference (in radians) between the applied angle and $\pi/2$. By this definition, leakage includes all photon-nonconserving transitions both within and out of the computational subspace. Qubit-specific, DAC-independent noise sources, preparation and readout errors, and the non-local effects of other DACs were omitted.\par

The Circuitizer results in Fig.~\ref{fig6_sim} reveal that there is a tradeoff between DAC under- and over-damping. Circuitizer incorporates as many quantum states as are necessary in order to achieve convergence of the dynamical evolution down to a specified error threshold. In this case, 20 quantum states were used when modeling the quantum components of the  device under test. As we decrease the resistance, we reduce the noise-free waveform PSD at the qubit frequency that drives leakage errors, while as we increase the resistance, we reduce resistive noise-induced phase errors. The physical mechanism for the latter is that noise with frequency comparable to the DAC hold time will shift the point in time where soliton injection (waveform rise) occurs relative to anti-soliton injection (waveform fall). Fig.~\ref{fig6_sim} shows that with increased damping compared to the DACs we designed, the average soliton DAC-induced leakage error is minimized when the slew rate is 175 ps or slower, consistent with the theoretical expectation that our waveform should be slower than the transmon $f_{01}^{-1} = 150$ ps, in order to avoid excitations. However, the average gate infidelity is minimized for DAC slew rates in the range of 60-115 ps, to around $4 \times 10^{-6}$, due to resistor noise-induced phase errors. The noise sampling method enables us to quantify both the trend in errors vs. DAC damping (illustrated by the moving-average), and the magnitude of the spread in error probability due to stochastic variation. \par

\section{Conclusion}
We designed a chip containing 30 soliton DACs that were controlled by an AWG, and benchmarked their phase gate fidelities on transmons. By performing Ramsey and qubit excitation/relaxation experiments with a varied number of DAC pulses, we determined that phase noise dominated our gate errors. Furthermore, nearly the same error was observed in ancilla qubits, that were not directly connected to the DACs, when the DAC's input drive amplitude exceeded the soliton injection threshold. In simulation, we identified the likely culprit of this noise as JJ superconducting phase oscillations within the middle of the DACs. Our simulations show that, by adding modest resistive shunts to each JJ in the middle of the DAC, we can damp out these oscillations without introducing an appreciable relative delay between the input and output DAC waveform. As these non-local EMI are mitigated, the gate fidelity is expected to approach the simulated single-DAC fidelity. In simulation, the fidelity can be optimized by slowing down the DAC waveform slew rate until it's comparable to the qubit $f_{01}$ frequency, achieving a minimum ensemble-averaged gate error of $4 \times 10^{-6}$. This is well below the surface code threshold, before accounting for separate effects such as intrinsic qubit noise and non-local DAC-induced errors.\cite{fowler2012surface, PhysRevA.80.052312}.

We have also demonstrated that the soliton DAC provides excellent protection from control noise. When a transmon was prepared in a superposition state, the soliton DAC completely buffered the transmon from passive phase errors up until the applied control noise reached the soliton injection threshold, $\mathrm{I_{\mathrm{noise}}} > \mathrm{I_{\mathrm{crit}}}$. On the other hand, when we applied below-threshold voltage noise noise to the transmon directly, it completely destroyed the qubit $T_2$ time. Since $\mathrm{I_{\mathrm{crit}}}$ is the drive amplitude threshold needed to perform a phase gate, this demonstrates that the DAC turns a noisy classical control pulse at its input into an analog-tunable, smooth, low-excitation, high signal-to-noise waveform for quantum applications.

\section*{Acknowledgments}
Tektronix is a registered trademark of Tektronix, Inc. The authors thank Twymun Safford, Jordan Sills, Moe Khalil, Joel Strand, Bradley Christensen, Patrick Warner, Joonbum Kwon, and Anthony Przybysz for support. This work was supported in part by the United States Government.

\bibliography{ref}
\end{document}